\documentclass[twocolumn]{IEEEtran}

\usepackage{epsfig}

\usepackage{}

\def\BibTeX{{\rm B\kern-.05em{\sc i\kern-.025em b}\kern-.08em
    T\kern-.1667em\lower.7ex\hbox{E}\kern-.125emX}}

\begin{document}
\title{Soft Handoff and Uplink Capacity in a Two-Tier CDMA System}
\author{Shalinee Kishore, Larry J. Greenstein, H. Vincent Poor, Stuart C. Schwartz\thanks{S. Kishore is with the Department of Electrical
and Computer Engineering, Lehigh University, Bethlehem, PA, USA.
H.V. Poor and S.C. Schwartz are with the Department of Electrical
Engineering, Princeton University, Princeton, NJ, USA.
L.J. Greenstein is with WINLAB, Rutgers University, Piscataway, NJ, USA.  This research was jointly supported
by the New Jersey Commission on Science and Technology, the
National Science Foundation under Grant CCR-00-86017, and the
AT\&T Labs Fellowship Program. Contact:
skishore@eecs.lehigh.edu.}}
\markboth{IEEE Transactions on Wireless Communications}{Kishore, et al: Soft Handoff and Uplink Capacity in a Two-Tier CDMA System}

\maketitle
\begin{abstract}
This paper examines the effect of soft handoff on the uplink user capacity of a CDMA system consisting of a single macrocell in which a single hotspot microcell is embedded.  The users of these two base stations operate over the same frequency band.  In the soft handoff scenario studied here, both macrocell and microcell base stations serve each system user and the two received copies of a desired user's signal are summed using maximal ratio combining.  Exact and approximate analytical methods are developed to compute uplink user capacity.  Simulation results demonstrate a 20\% increase in user capacity compared to hard handoff.  In addition, simple, approximate methods are presented for estimating soft handoff capacity and are shown to be quite accurate.
\end{abstract}
\section{Introduction}
In systems using code-division multiple access (CDMA), user capacity is enhanced by using {\em soft handoff}, \cite{viterbisho}-\cite{wongsho}.  By soft handoff we mean that a given user communicates simultaneously with two or more base stations until its path gain to just one of them is several dB stronger than its path gain to any other.  This is in contrast to {\em hard handoff}, wherein the user communicates at any given time with only one base, selected according to some path gain criterion.
\\
\\
In \cite{skishore1}, we examined a CDMA system consisting of a single macrocell in
which a single {\em hotspot} microcell is embedded.  Such a microcell
might be installed to serve small regions of high user density with a
low-cost, low-power base station.  Using both analysis and simulation,
we determined the uplink user capacity (i.e., the number of uplink
users supported with a specified probability of success) of this
two-tier system under the condition of hard handoff.  Here, we address
the same system when both bases serve each user (soft handoff), and we
quantify the gain in uplink user capacity.  The gain is relative to
the hard handoff approach, wherein the base selected is the one having
the largest path gain to the user terminal.  We focus on the uplink
direction because it tends to be the limiting direction, as we showed
in \cite{downlink}.
\\
\\
Section \ref{assump} describes the system geometry and the underlying
propagation and processing assumptions.  Section \ref{shocap} presents an exact criterion for determining uplink user capacity in ideal soft handoff, and compares the results to those for hard handoff \cite{skishore1}.  Also presented are two simpler methods, based on an analytical approximation, and they are shown to be quite accurate.  While our study is specific to a particular two-tier system in a highly dispersive multipath environment, extensions of our methods to other cases are possible, as we discuss in Section \ref{conc}.
\section{Assumptions}
\label{assump}
As in \cite{skishore1}, we assume a coverage region $\mathcal R$ (Figure 1) with a
macrocell base at the origin of the coordinate system and a microcell
base at a distance $x_0$ along the $x$ axis.  For concreteness in our
computations, we assume the shape of $\mathcal R$ is a square of side $S$,
with the macrocell base at the center.  It was shown in \cite{thesis1}, Ch. 4,
that the uplink capacity for square and circular $\mathcal R$ are
virtually the same; here, the square shape is assumed in order to
simplify computations.
\\
\\
All system users desire a rate $R$ and use a processing gain $W/R$,
where $W$ is the system bandwidth.  We assume that the potential users are made up of two populations: One consists of low-density (LD) users, distributed uniformly over the entire coverage region; the other consists of high-density (HD) users, distributed uniformly over a small square area surrounding the microcell base.  The side of this smaller square is $s \ll S$.  We denote by $P_h$ the probability that a randomly selected user is from the HD population.  The probability that a user is from the LD population is, of course, $1-P_h$.  We call $P_h$ the {\em hotspot density}.
\\
\\
The path gain, $T$, between either base and a user at a distance $d$ is assumed to be
\begin{eqnarray}
T = \left\{ \begin{array}{cc} H \left( \frac{b}{d} \right)^2 10^{\chi/10}, & d \leq b \\ H \left( \frac{b}{d} \right)^4 10^{\chi/10}, & d>b \end{array} \right. ,
\end{eqnarray}
where $b$ is the ``breakpoint distance'' \cite{rappaport} (in the same units as $d$) at
which the slope of the dB path gain versus distance changes; $\chi$
is a zero-mean Gaussian random variable for each user distance, with standard deviation $\sigma$; and $H$ is a proportionality constant that depends on wavelength, antenna heights, and antenna gains.  Note that $T$ is a local spatial average, so that multipath effects are averaged out.  There can be different values of $b$ for the microcell and macrocell, and similarly for $\sigma$ and $H$.  The factor $10^{\chi/10}$ is often referred to as lognormal shadow fading, which varies slowly over the terrain.  Both $\chi$ and $d$ are random variables for a randomly selected user.
\\
\\
In analyzing soft handoff performance, we will make the following
assumptions:  (1) Each user in the environment is processed by {\em
both} bases, regardless of the user's path gains to the two
bases.\footnote{This is an idealized assumption, since in practical
systems only a subset of users (usually those on or near cell
boundaries) is engaged in soft handoff.}  (2) At each base, ideal
RAKE processing is performed on each user's received signal in order
to maximize diversity;  (3) Each user-base path is ``infinitely dispersive,'', meaning that there is a near-infinitude of resolvable, comparable-strength multipaths on each link.  Given (2) and (3), the receiver output signal sample for each user will be non-fading, just as if the link had a single path of fixed gain.  We can therefore proceed by assuming the additive white Gaussian noise (AWGN) condition for each link, which simplifies the analysis.  The implications of this assumption, and its relaxation, are discussed in Section IV. 
\section{Analysis and Results}
\label{shocap}
\subsection{Exact Method}
From the foregoing, we can proceed as though user $i$ ($i=1,2,\ldots N$) has signal path gains $\sqrt{T_{Mi}}e^{j\phi_{Mi}}$ and $\sqrt{T_{\mu i}}e^{j\phi_{\mu i}}$ to the macrocell and microcell, respectively.  This user transmits a signal at power $P_i$, which is controlled by the bases. The parameters $\phi_{Mi}$ and $\phi_{\mu i}$ are
the phases of the path gains from user $i$ to the macrocell and
microcell bases, respectively.  In the soft-handoff scenario
studied here, both the macrocell and microcell bases receive,
despread, and RAKE-combine the signal from each of the $N$
users.  There are, therefore, two output streams (one at the
macrocell and one at the microcell) which contain a user's desired
signal.  Each of these two streams also contains interference
from the $N-1$ other users plus thermal noise. 
\\
\\
The output streams
at the macrocell and microcell bases are weighted by $w_{Mi}$ and
$w_{\mu i}$, respectively, and then summed. After combining, the overall output SINR (signal-to-interference-plus-noise ratio) of user $i$ is
\begin{equation}
SINR_i = \frac{\frac{W}{R} S_i}{| w_{Mi}|^2 I'_{Mi} +|w_{\mu i}|^2 I'_{\mu i}}, \label{sho_sinr_eq}
\end{equation}
where 
\begin{eqnarray}
S_i &=& \left| w_{Mi} e^{j \phi_{Mi}}\sqrt{P_i
T_{Mi}} + w_{\mu i} e^{j \phi_{\mu i}} \sqrt{P_i T_{\mu i}}
\right|^2, \\
I'_{Mi} &=& \left( \sum_{j=1,j \neq i}^N P_j T_{Mi}
+\eta W \right), \\
I'_{\mu i} &=& \left( \sum_{j=1, j \neq i}^N P_j
T_{\mu i} +\eta W \right),
\end{eqnarray}
and $\eta W$ is the thermal noise power in the system bandwidth.  This SINR equation implicitly treats the interference from each user $j$ as if it is uncorrelated between bases.  In fact, the two interferences from user $j$ {\em are} correlated, as they carry the same data.  The combiner output voltage sample from user $i$ is therefore proportional to
\begin{eqnarray*}
w_{Mi} \sqrt{P_i T_{Mi}}e^{j \phi_{Mi}} + w_{\mu i}\sqrt{P_i T_{\mu i}} e^{j \phi_{\mu i}}
\end{eqnarray*}
and the squared magnitude has a cross-term proportional to the cosine of a path-gain-related phase.  The phase term is uniformly random on $[0,2\pi)$, and so the sum over $j \neq i$ of all cross-terms will converge to a mean of zero as $N$ approaches infinity.  For the values of $N$ of interest here, substitution of this mean value is a very reasonable approximation; hence (2) applies.
\\
\\
The system studied here uses path-gain-weighted combining, i.e., it co-phases the two
signals, via the phases of $w_{Mi}$ and $w_{\mu i}$, and uses the
following magnitudes for $w_{Mi}$ and $w_{\mu i}$:
\begin{equation}
|w_{Mi}|= \frac{\sqrt{T_{Mi}}}{\sqrt{T_{Mi}+T_{\mu i}}}\mbox{~~~and~~~}
|w_{\mu i}|= \frac{\sqrt{T_{\mu i}}}{\sqrt{T_{\mu i}+T_{\mu i}}}.
\end{equation}
Observe that $w_{Mi}^2+w_{\mu i}^2=1$.  Using these values and the
assumed co-phasing, we can simplify (2):
\begin{equation}
SINR_i = \frac{\frac{W}{R} P_i(T_{Mi}+T_{\mu
i})}{|w_{Mi}|^2 \sum_{j \neq i}^N P_j
T_{Mj}+|w_{\mu i}|^2 \sum_{j \neq i}^N
P_j T_{\mu j} + \eta W}. \label{sho_sinr_new}
\end{equation}
We desire $SINR_i \geq \Gamma$, where $\Gamma$ is the minimum required
SINR for each user.  Setting $SINR_i = \Gamma$ for all $i$ leads to a matrix solution to the $P_i$'s that has the following form:
\begin{equation}
\mathbf{P}=\frac{\eta W}{K'} \mathbf{A}^{-1}\mathbf{1}, \label{sho_pwrs}
\end{equation}
where $K'=W/(R \Gamma)$ (note $K'+1$ is the {\em single-cell pole
capacity} \cite{gilhousen});
$\mathbf{P}=[P_1\mbox{~~}P_2\mbox{~~}...\mbox{~~}P_N]^T$; $\mathbf{A}$ is an $N$ x $N$ matrix, with
\begin{eqnarray}
A_{ii} &=& T_{Mi}+T_{\mu i} \mbox{~~~and~~~} \\
A_{ij} &=& -\frac{1}{K'}\frac{T_{Mi}T_{Mj}+T_{\mu i}T_{\mu j}}{T_{Mi}+T_{\mu i}}, \mbox{~~~}i \neq j;
\end{eqnarray}
and $\mathbf{1}$ is an $N$ x $1$ vector with each element equal to
one.  Note that the elements of $\mathbf{A}$'s are random variables, since each path gain $T_{Mi}$ ($T_{\mu i}$) is determined by the random location and shadow fading of user $i$ relative to the macrocell (microcell).  
\\
\\
The two-tier system can support $N$ total users if and only if the
transmit power levels in (\ref{sho_pwrs}) are positive, and this occurs if
and only if the determinant of $\mathbf{A}$ is positive.  The
randomness of $\mathbf{A}$ implies that this {\em feasibility} of
supporting $N$ users occurs with some probability.  The size of the
random matrix $\mathbf{A}$ complicates the
actual calculation of this probability of feasibility.  Nevertheless, it is possible to use simulation methods to
perform this calculation and gain insight into the capacity
performance of the two-tier system under soft-handoff.
\\
\\
We performed simulations to find the 95\% value of
$N$ using the following method:  For the two-tier, two-cell system, with parameters as in Table 1, we
performed 10,000 trials.  In each trial, the two path gains for each of
$N$ randomly-selected users were
generated.  These gains were then used to form the
matrix $\mathbf{A}$ and compute its determinant.  The feasibility
(success) of the user set was determined by the sign of this determinant, and the probability of success over the 10,000 trials was computed for the selected $N$.  The
simulation program obtained the value of $N$ for which 95\% of trials
yielded feasibility.  The results are presented in Figure
\ref{sho_results1}, where we plot $N$ as a function of $P_h$, the
hotspot density. We also plot the
hard-handoff results from \cite{skishore1}.   The plots show that
user capacity varies with $P_h$, the maximum occurring at roughly
$P_h=0.5$.  This is the density for which, on average, half of all
users lies in the microcell coverage area.  With this condition, the
system contains roughly an equal number of users per base, which our
previous work shows leads to maximum capacity \cite{thesis1}, Chap. 4.   Figure
\ref{sho_results1} also shows that for all $P_h$, the capacity is higher for soft handoff than for hard-handoff, and
these gains are at most 20\%.

\subsection{Approximate Method}
\label{approx1}
Due to the complexity in computing the exact user capacity for
soft handoff (as outlined above), we now present an approximation method.  We resort to the fiction that, {\em in the denominator} of (2), $|w_{Mi}|^2$ and
$|w_{\mu i}|^2$ are the same for all users; we call these weights $|w_M|^2$ and
$|w_\mu|^2$ and the common values are assumed to be
\begin{equation}
|w_{M}|^2 = \frac{1}{N} \sum_{i=1}^N |w_{Mi}|^2 = \frac{1}{N}
\sum_{i=1}^N \frac{T_{Mi}}{T_{Mi}+T_{\mu i}} \label{sho_wm}
\end{equation}
and
\begin{equation}
|w_{\mu}|^2 = \frac{1}{N} \sum_{i=1}^N |w_{\mu i}|^2 =
\frac{1}{N}\sum_{i=1}^N \frac{T_{\mu i}}{T_{Mi}+T_{\mu i}}.
\label{sho_wmu}
\end{equation}
In other words, the weights $|w_M|^2$ and $|w_{\mu}|^2$ are the
average values of $|w_{Mi}|^2$ and $|w_{\mu i}|^2$ over the $N$
users. Using these constant weights, the
denominator in (\ref{sho_sinr_eq}) is
\begin{eqnarray*}
D_i = |w_M|^2 \sum_{j=1}^N P_j T_{Mj} + |w_\mu|^2 \sum_{j=1}^N P_j T_{\mu j} \\ \qquad -|w_M|^2P_iT_{Mi} - |w_\mu|^2P_iT_{\mu i}+\eta W.
\end{eqnarray*}
This term is independent of $i$ except for the third and fourth
terms, which are small compared to the first and second terms for $N$ large.  If
we ignore them, we obtain a common ($i$-independent) denominator
given by
\begin{equation}
D=|w_M|^2 \sum_{j=1}^N P_j T_{Mj} + |w_\mu|^2 \sum_{j=1}^N P_j T_{\mu j}+\eta W
\label{constantD}
\end{equation}
Assuming the denominator in (\ref{sho_sinr_new})
is replaced by $D$, and setting $\mbox{SINR}_i$ to the required value of
$\Gamma$, we obtain
\begin{equation}
P_i=\frac{1}{T_{Mi}+T_{\mu i}}\cdot \frac{D}{K'}.
\end{equation}
Substituting
this into (\ref{constantD}), we obtain
\begin{eqnarray*}
D = \left[ |w_M|^2 \sum_{j=1}^N\frac{T_{Mj}}{T_{Mj}+T_{\mu j}} +
|w_\mu|^2 \sum_{j=1}^N \frac{T_{\mu j}}{T_{\mu j}+T_{Mj}} \right]
\frac{D}{K'} \\
+ \eta W.\qquad
\end{eqnarray*}
Solving for $D$, we get
\begin{equation}
D=\frac{\eta WK'}{K'-\left[|w_M|^2
\sum_{j=1}^N\frac{T_{Mj}}{T_{Mj}+T_{\mu j}} + |w_\mu|^2
\sum_{j=1}^N \frac{T_{\mu j}}{T_{\mu j}+T_{Mj}} \right]}.
\end{equation}
Using (\ref{sho_wm})-(\ref{sho_wmu}) for $|w_M|^2$ and
$|w_\mu|^2$, we see that this quantity (i.e., each $P_i$) is
positive if and only if
\begin{equation}
\left( \sum_{i=1}^N \frac{T_{Mi}}{T_{Mi}+T_{\mu i}} \right)^2 + \left(
\sum_{i=1}^N \frac{T_{\mu i}}{T_{Mi}+T_{\mu i}} \right)^2 < N K'.
\label{sho_feas_approx}
\end{equation}
This feasibility condition is far-simpler to examine than
$det(\mathbf{A})>0$. 
\\
\\
We applied this approximation technique to the same system studied above.
  As before, we used 10,000 trials for each of several values of $N$.
In each trial, we generated random user locations and shadow fadings
for $N$ users, leading to $N$ pairs of $(T_{Mi},T_{\mu i})$.  Using
(\ref{sho_feas_approx}), we then determined whether power control was feasible.  Doing
this for all trials, we were able to compute a probability of
feasibility for the given $N$.  Finally, we found the value of $N$
corresponding to 95\% probability.  This $N$ is plotted against $P_h$
in Figure 2 (``Approx. 1'').  We see that
there is very good agreement; the approximation over-estimates capacity by
at most three users, or roughly 7\%.  

\subsection{Alternate Approximation Method}
We now develop an even simpler analytical method to compute uplink capacity. We first note that (14) can be rewritten as
\begin{equation}
|A + jB|^2 < NK',
\label{new17}
\end{equation}
where 
\begin{eqnarray}
A &=& \sum_{i=1}^N \frac{T_{Mi}}{T_{Mi}+T_{\mu i}} \mbox{~~~and~~~} \\
B &=& \sum_{i=1}^N \frac{T_{\mu i}}{T_{Mi}+T_{\mu i}}=N-A.
\end{eqnarray}
Since $A$ and $B$ are random variables, we can rewrite them as $A =
\mbox{E} \left\{ A \right\} + \delta (A)$ and $B = \mbox{E} \left\{
B \right\} + \delta (B)$, respectively, where  $\delta (A)$ and $\delta(B)$ represent
random perturbations of $A$ and $B$ about their mean values,
$\mbox{E}\left\{ A \right\}$ and $\mbox{E}\left\{ B \right\}$.  Based
on this new representation, (\ref{new17}) changes to 
\begin{equation}
|X|^2 < N K', 
\label{newnew17}
\end{equation}
where $X = |C + \delta (A) + j \delta (B)|$ and $C=\mbox{E} \left\{ A \right\} + j \mbox{E}\left\{ B
\right\}$.  Since $B=N-A$, we have, $\delta (B) = - \delta
(A)$.  We make this substitution and put the complex value $X = C + \delta (A) - j\delta (A)$ on a
better basis by performing phase rotation.  This is done by
multiplying $X$ by $(\mbox{E}\left\{ A \right\}-j
\mbox{E}\left\{ B \right\})/|C|$.  The rotated representation of $X$ is then
\begin{equation}
X' = |C|+\frac{(2 \mbox{E}\left\{ A \right\}- N)\delta (A)}{|C|}-j\frac{N \delta (A)}{|C|},
\end{equation}
where we have used $B=N-A$.  The feasibility condition in (\ref{newnew17}) can now be written as
$|X'|^2 < NK'$.
\\
\\
A key step in this approximation is to first assume that $|X'| \approx
|Re(X')|$.  That is, assuming $|C|^2$ is large with respect to the
standard deviation of $\delta (A)$ (as it is for large $N$), we ignore the quadrature term
of $X'$.  Thus, the feasibility condition is approximated as
\begin{equation}
|C|+\frac{(2 \mbox{E}\left\{ A \right\} - N)\delta (A)}{|C|} < \sqrt{N K'}
\label{newnewnew17}
\end{equation}
Next, we use the central limit theorem to approximate $\delta (A)$, which 
is a sum of i.i.d. random terms, as a zero-mean Guassian variate.  Let us represent the left side of (\ref{newnewnew17}) as $Z$, with mean $\mu_Z$ and standard deviation $\sigma_Z$.  If $\mu$ and $\sigma^2$ are the mean and variance of
each i.i.d. term in the sum $A$, (16), then it can be
shown that 
\begin{eqnarray}
\mu_Z &=& |C| = N \sqrt{1-2 \mu+2 \mu^2} \mbox{~~~and~~~}\\
\sigma_Z &=& \frac{N^{3/2} \sigma |2 \mu-1|}{|C|}.
\end{eqnarray}
This implies that feasibility occurs with probability $\alpha$ or greater if and only if
\begin{equation}
N \leq \left( \frac{\sqrt{K'} - \frac{u_\alpha \sigma |2
\mu - 1 |}{\sqrt{1-2\mu +2 \mu^2}}}{\sqrt{1-2
\mu+2 \mu^2}}
\right)^2,
\label{brandnew}
\end{equation}
where $u_\alpha$ is the $\alpha$-th percentile value of a zero-mean,
unit-variance Guassian variate.  This approximation produces a simple,
closed-form solution for the capacity supported in a two-tier system for a desired level of feasibility,
$\alpha$.  Given $\alpha$ (assumed here to be 0.95) and values of $\mu$ and
$\sigma^2$, which can be determined using either simulation or
analysis, user capacity can be simply estimated. 
\\
\\   
We used simulation to
estimate $\mu$ and $\sigma^2$ for numerous values of $P_h$. Then, invoking
(\ref{brandnew}) and rounding to the next lower integer, we obtained the result shown in Figure 2
(``Approx. 2'').  We see that this new approximation overestimates the results from Section \ref{approx1} by at most one user and exceeds the results from simulation by at most 4
users, corresponding to 10\%.  
\section{Discussion and Conclusion}
\label{conc}
We have devised three methods of soft handoff assessment that become
progressively simpler to implement, in exchange for modest
overestimations of capacity (less than 10\%). Our results suggest that the uplink capacity benefit of soft
handoff over hard handoff is minor for the two-tier, two-cell
system we have examined.  As shown in Fig. 2, the gain is no more than
20\% for any value of hotspot density.  However, this gain will be
higher if the microcell base transmits and receives at a higher power,
i.e, if $H_\mu$ is larger.  With a larger $H_\mu$, each user's total
received power after path-gain-weighted combining will be higher,
whereas under hard handoff, the attainable user capacity does not change
as $H_\mu$ increases \cite{thesis1}.  The gain due to
soft handoff will also be higher if this mechanism is used in a system with multiple
embedded microcells.  Each embedded microcell offers an additional
path from the user and thus improves the total received power after
path-gain-weighted combining.  The techniques presented here can be
used to determine the capacity gains when $H_\mu$ increases; the gain
offered by multiple embedded microcells is a topic of future research.
\\
\\
Our results are obtained for highly dispersive channels and ideal RAKE
receivers.  For scenarios exhibiting less protection against multipath
\cite{skishore_globe1}, soft handoff provides a micro-diversity
benefit (against multipath fading) and a macro-diversity
benefit (against shadow fading), so the soft handoff improvement will
be greater.  The key point, however, is that, as CDMA systems become
more and more wideband, the capacity improvement due to soft
handoff will become more modest. 
\bibliographystyle{IEEEbib}

\begin{thebibliography}{1}
\bibitem{viterbisho}
A.~Viterbi, et. al.,
\newblock ``Soft Handoff Extends CDMA Cell Coverage and Increases Reverse Link Capacity,''
\newblock {\em IEEE Journal on Selected Areas in Communications,} Vol. 12, No. 8, pp. 1281--87, October 1994.
\bibitem{patelsho}
P. Patel, et. al.,
\newblock ``A Simple Analysis of CDMA Soft Handoff Gain and its Effect on the Cell's Coverage Area,'' 
\newblock {\em Wireless Information Networks}, J. Holtzman, editor, pp. 155--172, Kluwer Academic, Boston, MA, 1996.
\bibitem{steelesho}
C.C. Lee and R. Steele,
\newblock ``Effect of Soft and Softer Handoffs on CDMA System Capacity,'' 
\newblock {\em IEEE Transactions on Vehicular Technology}, Vol. 47, Issue 3, pp. 830--841, August 1998.
\bibitem{shapira}
Shapira, J.,
\newblock ``Microcell Engineering in CDMA Cellular Networks,''
\newblock {\em IEEE Transactions on Vehicular Technology}, Vol. 4, Issue 4, pp. 817--825, August 1994.
\bibitem{wongsho}
Wong, D. and T.J. Lim,
\newblock ``Soft Handoffs in CDMA Mobile Systems,''
\newblock {\em IEEE Personal Communications}, Vol. 4, Issue 6, pp. 6--17, December 1997.
\bibitem{skishore1}
S.~Kishore, et. al.,
\newblock ``Uplink Capacity in a CDMA Macrocell with a Hotspot Microcell: Exact
  and Approximate Analyses,"
\newblock {\em IEEE Transactions on Wireless Communications}, Vol. 2, No. 2, pp. 364-374, March 2003.
\bibitem{downlink}
S. Kishore, et. al.,
\newblock ``Downlink User Capacity in a CDMA Macrocell with an
Embedded Hotspot Microcell,'' in {\em Proceedings of Globecom},
vol. 3, pp. 1573-1577, December 2003.
\bibitem{thesis1}
S.~Kishore, 
\newblock ``Capacity and Coverage of Two-Tier Cellular CDMA Networks,''
\newblock {\em Ph.D. Thesis}, Department of Electrical Engineering, Princeton University, January 2003.
\bibitem{rappaport}
T. Rappaport,
\newblock {\em Wireless Communications: Principles and Practice},
2nd Edition, Prentice Hall, Englewood Cliffs, NJ, 2002.
\bibitem{gilhousen}
K.S. Gilhousen, et al.,
\newblock ``On the Capacity of a Cellular CDMA System,''
\newblock {\em IEEE Transactions on Vehicular Technology}, Vol. 40, pp. 303-312, May 1991.
\bibitem{skishore_globe1}
S. Kishore, et al.,
\newblock ``User Capacity in a CDMA Macrocell with a Hotspot Microcell:  Effects of Transmit Power Constraints and Finite Dispersion,''
\newblock in {\em Proceddings of Globecom}, vol. 3, pp. 1558-1562, December 2003.
\end{thebibliography}

\vspace{.5in}
\begin{table}[h]
\begin{center}
\begin{tabular}{|c|c||c|c|}
\hline $W/R$ & 128 & -- & --
\\ $\Gamma_M$ & 7 dB & $\Gamma_\mu$ & 7 dB
\\ $b_M$ & 100 m & $b_\mu$ & 100 m
\\ $H_M$ & $10 H_\mu$ & $x_0$ & 300 m
\\ $\sigma_M$ & 8 dB & $\sigma_\mu$ & 4 dB
\\ $s$ & 200 m & $S$ & 1 km \\
\hline
\end{tabular}
\caption{System parameters used.}
\end{center}
\label{sys_param}
\end{table}
\vspace{.5in}
\begin{figure}[h]
\begin{center}
\epsfig{figure=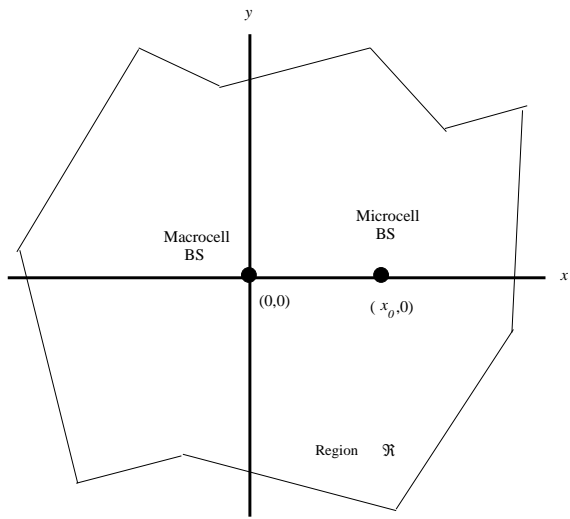,width=3in} \caption{A general description
of the coverage area $\mathcal R$ containing one macrocell base station
and one microcell base station. Results presented here are relatively insenstive to $x_0$.} \label{sysfig}
\end{center}
\end{figure}
\begin{figure}[h]
\begin{center}
\epsfig{figure=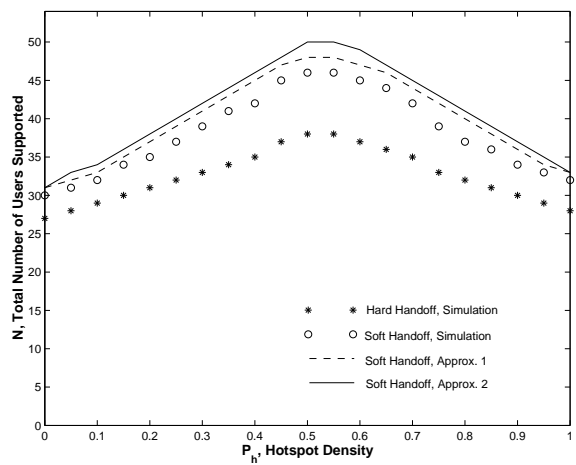,width=3in} \caption{Uplink user capacity versus hotspot density.}
\label{sho_results1}
\end{center}
\end{figure}

\end{document}